\documentclass[a4paper,11pt]{article}
\usepackage{pos}
\newcommand*{\tran}{^{\mkern-1.5mu\mathsf{T}}}
\usepackage{amsmath}
\DeclareMathOperator{\Tr}{Tr}

\title{Trivializing Flow in 2D O(3) sigma model}

\author*[a]{Christopher Chamness}
\author[a,b]{Konstantinos Orginos}
\author[a]{Daniel Kovner}

\affiliation[a]{Department of Physics, William \& Mary,\\
  300 Ukrop Way, Williamsburg, VA 23185}

\affiliation[b]{Thomas Jefferson National Accelerator Facility\\
12000 Jefferson Ave, Newport News, VA 23606}

\emailAdd{cdchamness@wm.edu}
\emailAdd{kostas@wm.edu}
\emailAdd{dkovner@wm.edu}

\abstract{
The two dimensional O(3) sigma model, just as quantum chromodynamics,  is an asymptotically free theory with a mass gap. Therefore, it is an interesting and simple toy model to investigate algorithms for Markov Chain Monte Carlo simulations of quantum chromodynamics. In this talk, we discuss the construction of a trivializing map, a field transformation from a given theory to a trivial one, through a suitably chosen gradient flow. 
An analytic solution for the generating functional of this trivializing flow has been obtained by
a perturbative expansion in the flow time. Utilizing this solution allows for new approaches
to be considered when proposing updates for a Markov Chain  Monte Carlo algorithm.}

\FullConference{The 40th International Symposium on Lattice Field Theory (Lattice 2023)\\
July 31st - August 4th, 2023\\
Fermi National Accelerator Laboratory\\}

\begin{document}
JLAB-THY-24-3981
\maketitle

\section{Introduction}
The idea of using a Trivializing Map for improving Markov Chain Monte Carlo (MCMC) calculations in Quantum Chromodynamics (QCD) has been proposed more than 10 years ago~\cite{Luscher:2009eq}. In this approach, the field variables that satisfy a non-trivial distribution  are mapped to new variables that satisfy a trivial distribution.
This is done by via a bijective map that transforms the trivial probability density distributions to the target one.
Although, a very promising idea, up to now it has not found an efficient application to QCD. In this work, we turn our attention to a much simpler model, the O(3) sigma model in two dimensions. This model, shares with QCD the properties of asymptotic freedom~\cite{POLYAKOV197579} as well as the existence of mass gap~\cite{Hasenfratz:1990zz}.

It is, therefore, a non-trivial theory that might help us  understand the properties of trivializing maps that could be relevant to QCD. Furthermore, this toy model is  less computationally intensive and allows for new algorithms to be designed and tested much quicker than would be possible in QCD~\cite{Bietenholz:2018agd}. A similarly motivated study has already been made in the case of the CP(N-1) model in two dimensions~\cite{Engel:2011re}, where the leading order approximation in the flow time expansion was used to study the potential of improvement of the corresponding Hybrid Monte Carlo algorithm. In this contribution, we discuss the perturbative flow time expansion of the generating functional of the gradient flow which defines our trivializing map.

\section{Trivializing Map}
A Trivializing Map~\cite{Luscher:2009eq} may be defined as a map between the initial condition and the solution of a gradient flow at a specific flow time t. Without loss of generality, this time is taken to be $t=1$. This map is then used to transform random variables with a non-trivial distribution to random variables with a trivial distribution (i.e. easy to sample from). 
 The gradient flow is defined by a  generating functional, $\tilde S[s]$, that is a scalar function of the field variables, $s(x)$. The generator of the flow is defined through its gradient.
\begin{equation} \label{sDotDef}
    \dot s(x) = \partial_x \tilde S[s] \quad \quad \partial_x \equiv \frac{\partial}{\partial s(x)}
\end{equation}
Given that the initial conditions of the fields, $s(x, 0) = \sigma$ at $t =0$, the map is then defined as 
\begin{equation}
    s(x,t) = \mathcal{F}^{-1}_t(\sigma)
\end{equation}
The field variables, $s(x, t)$, are time-dependent values that are determined from the initial conditions. The map, $\mathcal{F}^{-1}_t$, due to the uniqueness of solutions of the flow equation, is bijective and is therefore invertible, $\mathcal{F}_t(s(x, t)) = \mathcal{F}_t(\mathcal{F}_t^{-1}(\sigma)) = \sigma$. This map can be seen as a change of variables between the distributions of the $s$ and $\sigma$ fields. 
\begin{equation}
    q(\sigma)\mathcal{D}\sigma= q(\mathcal{F}_1(s))\det\mathcal{J}(s)\mathcal{D}s
\end{equation}
where $\mathcal{J}$ is the Jacobian of the transformation. This is the trivializing map if it transforms the trivial distribution into the target distribution.
\begin{equation}
    p(s)\mathcal{D}s = q(\mathcal{F}_1(s))\det\mathcal{J}(s)\mathcal{D}s
\end{equation}
Taking the log of both sides results in
\begin{equation} \label{SimpleTrivFlow}
    S_{t}(s) = S_0 (\mathcal{F}_1(s)) - \ln[\det \mathcal{J}(s)] - C\,,
\end{equation}
where $S_{t}$ and $S_0$ are the target and trivial actions respectfully and $C$ is a constant determined by the partition function of the distributions. To go further, we must consider the Jacobian, $\mathcal{J}$, for the specific change of variables we are interested in. In our case the 2D-O(3) sigma model.

\section{The 2D-O(3) sigma model}
The lattice 2D-O(3) sigma model is defined by the action, $\mathcal{S}[s]$, and the probability distribution, $p[s]$, where the integration measure, ${\cal D}s$ is over the unit sphere ${\cal S}^2$

\begin{equation}
    \mathcal{S}[s] = \frac{\beta}{2} \sum_x \sum_\mu ( 1 - s_x \cdot s_{x+\mu} ), \quad s_x \in \mathcal{S}^2, \quad x \in \mathbb{Z}^2, \quad \mu \in \{ \mathbb{Z}^2 \;\big|\; |\mu| = 1 \}
\end{equation}
\begin{align}
    p[s] = \frac{ e^{-\mathcal{S}[s]}}{\cal Z}, \quad \quad\quad \quad\quad \quad 
     {\cal Z} = \int {\cal D} s \, e^{-{\cal S} [s] }
\end{align}
To define a Trivializing Map from a gradient flow we need to first consider the complications that arise from the compact nature of the sigma model. The map has to transform the sphere to a sphere i.e. $\mathcal S^2 \rightarrow \mathcal S^2$. This implies that any transformation of the spin degrees of freedom must be an SO(3) rotation:
\begin{equation}
    \forall R \in SO(3)\quad \& \quad \forall s \in {\cal S}^2, \quad R*s \in {\cal S}^2
    \end{equation}
Therefore, the most general map that takes an $\mathcal{S}^{2}$ valued vector field $\sigma$ and maps to an $\mathcal{S}^{2}$ valued vector field $s$, can be written as
\begin{equation}
     s(x) = R(x)\, \sigma(x) \,,
\end{equation}
where $R(x)$ is an element of SO(3) which is determined up to an SO(2) rotation. Rotations about the axis of the vector are the identity transformation. This quotient subgroup SO(3)/SO(2) of SO(3) is diffeomorphic to the sphere ${\cal S}^2$ manifold.

It is useful to consider parametrizing the spin variables $s(x)$ as a rotation of a reference vector. This way the Lie derivative defined through an infinitesimal rotation can be expressed as following:
\begin{equation} 
     s(x) = R(x) s_0, \quad \quad \quad R(x) = \exp(\sum_\alpha \omega_x^\alpha L_\alpha)
\end{equation}
\begin{equation} \label{PartialDef}
     \partial_y^\alpha R(x) =  L_\alpha R(x)\; \delta_{x,y} \implies \partial_y^\alpha s(x) =  L_\alpha s(x)\,\delta_{x,y}\,.
\end{equation}
where $L_a$ are the traceless antisymmetric generators of $\mathfrak{so}(3)$.
This derivative satisfies the expected property
\begin{align}
    \sum_\alpha s^{a}(x) \partial^a_x s(x) = 0\,.
\end{align}
In other words, the derivatives are elements of the tangent bundle. Using a scalar functional of the fields, $\tilde S$, whose gradient defines the generator of the flow, we have a flow on the $S^2$ manifold as 
\begin{align} \label{O3sDotDef}
     \dot s(x,t) = -\sum_a L_a \partial_x^\alpha \tilde S [ s(x,t) ]\, s(x,t).
\end{align}
\section{Trivializing Flows in the Sigma Model}
With the fields and their derivatives defined, it is now clear what we mean by the Jacobian in Eq.~\eqref{SimpleTrivFlow}. The Jacobian Matrix is then given by:
\begin{align}
{\cal J}^{a,b}[R'(x)](x,y) = - \frac{1}{2}\Tr [\hat \partial_y^b R(x) R\tran(x) L_a]
\end{align}
where $\hat \partial_y^b$ is the Lie derivative with respect to the initial condition fields $\sigma(x)$.

In Eq.~\eqref{SimpleTrivFlow} we have written the target distribution as independent of flow time while, in principle, it could have explicit flow time dependence in which it interpolates between trivial and the final target distributions. The simplest case is a linear interpolation between the two distributions.
\begin{equation} \label{LinearIterAnsatz}
    S_t(s(t), t) = t [ S_0(s(t)) - \mathcal{S}(s(t))] + \mathcal{S}(s(t))
\end{equation}
Where at $t=0$, $S_t(s(t),t)$ is the target distribution, $\mathcal{S}$, and at $t=1$ it is the trivial distribution, $S_0$. Using this simplifying ansatz, we may take the total time derivative of our matching condition Eq.~\eqref{SimpleTrivFlow}. 
\begin{equation} 
    \sum_x \partial_x S_{t}(s) \dot s(t) + \partial_t S_{t}(s) = \frac{d}{dt} S_0 (\mathcal{F}_1(s)) - \frac{d}{dt} \ln[\det \mathcal{J}(s)] - \dot C
\end{equation}
From here, some simplifying steps may be taken. For any compact group, such as $SO(3)$, the trivial action is simply $S_0 = 0$, which is constant and thus has no time dependence. It can also be shown that the total time derivative of the Jacobian term may be simplified,
\begin{align}
\frac{d \ln\det [{\cal J}(R_t) ]}{dt} =\frac{d \Tr \ln[{\cal J}(R_t)]}{dt} = \Tr [ \dot {\cal J}(R_t) {\cal J}^{-1}(R_t)] = \partial_x^a \partial_x^a \tilde S
\end{align}
In which case, using Eqs.~\eqref{O3sDotDef} and~\eqref{LinearIterAnsatz} it can be shown that the matching condition reduces to:
\begin{equation}
    -\partial^2 \tilde S + t \sum_x \partial_x \mathcal{S} \; \partial_x \tilde S = - \mathcal{S} - \dot C
\end{equation}

Exact solutions for the generating functional, $\tilde S$, are not possible. However, the small flow time expansion lets us determine $\tilde S$ and $\dot C$ in a power series of $t$.
\begin{equation}
    \tilde S(s(t), t) = \sum_n t^n S^{(n)}(s(t)) \quad \quad \dot C(t) = \sum_n t^n \dot C^{(n)}
\end{equation}
Using this form and collecting like powers of $t$, we determine 2 classes of equations
\begin{align}
 n=0:&\;\;\;\;\;\;\;\;\;\;\;\;\;\; -\partial^2 S^{(0)} = -\mathcal{S} - \dot C^{(0)} \label{eq:order_by_order_0}\\
 n\geq 1:&\;\;\;\;\; -\partial^2 S^{(n)} + \sum_x \partial_x \mathcal{S}\; \partial_x S^{(n-1)} = - \dot C^{(n)}
 \label{eq:order_by_order_1}
\end{align}
These equations have no explicit dependence on $t$, and because they must hold for every $t$ we may consider these fields at any instant to have static values. Much like in the case of $SU(3)$ gauge theory~\cite{Luscher:2009eq}, this set of equations allows us to construct an iterative process to compute $S^{(n)}$ to any order, as each order depends only on our target action, $\mathcal{S}$, and the previous order solution. 

\subsection{Perturbative Solution }
The $0^{th}$ Order is a special case that only depends on the target action.
\begin{equation} \label{zerothOrder}
    -\partial^2 S^{(0)} = -\mathcal{S} - \dot C^{(0)},  \quad \quad \quad \quad \mathcal{S} = \frac{- \beta}{2} \sum_{x, \; \mu} s_x \cdot s_{x + \mu} 
\end{equation}
Consider the ansatz that $S^{(0)}$ has the same functional form as $\mathcal{S}$. 
\begin{equation}
    S^{(0)} = \gamma_0  \sum_{x, \; \mu} s_x \cdot s_{x + \mu} 
\end{equation}
Then, using Eqs.~\eqref{PartialDef} and~\eqref{zerothOrder} it is easy to show,
\begin{equation}
    -\partial^2 S^{(0)} = 4 S^{(0)} \implies 4 \gamma_0 = \frac{\beta}{2} \implies \gamma_0 = \frac{\beta}{8} \implies S^{(0)} = \frac{\beta}{8}  \sum_{x, \; \mu} s_x \cdot s_{x + \mu} 
\end{equation}

Using the  $0^{th}$ order solution and Eq.~\ref{eq:order_by_order_1} we can construct the $1^{st}$ order solution by noting that:
\begin{align} 
    \implies &-\partial^2 S^{(1)} = \frac{\beta^2}{4} \sum_{x} \Big[ \sum_{\mu,\nu} (s_{x} \cdot s_{x+\mu+\nu}) - (s_x \cdot s_{x+\mu})( s_x \cdot s_{x+\nu}) \Big]  -\dot C^{(1)} 
\end{align}

We may assume that $S^{(1)}$ has the same functional form as the RHS of the above equation. However, when applying $\partial^2$ this ansatz for $S^{(1)}$, a new type of term arises. So we must change the functional form to include the new term. This process is repeated until the set of terms in the functional form is closed under $\partial^2$. This generates a good basis for the solution space. Using this it is possible to solve for the $1^{st}$ order flow functional.

\begin{equation}
    S^{(1)} = \frac{\beta^2}{40} \Big[2 \Psi^{(2)} - \Tilde\Psi^{(1,1)} + \frac{1}{6} \Psi^{(1,1f)} \Big] 
\end{equation}
where,
\begin{equation}
   \Psi^{(2)} = \sum_{x} \sum_{\mu,\nu} s_{x} \cdot s_{x+\mu+\nu},\;
        \Tilde\Psi^{(1,1)} =
        \sum_{x} \sum_{\mu,\nu} (s_x \cdot s_{x+\mu})(s_x \cdot s_{x+\nu}), \; \Psi^{(1,1f)} = \sum_{x} \sum_{\mu} (s_{x} \cdot s_{x+\mu})^2
\end{equation}

Following the same steps as the 1st Order, continually adding terms to the ansatz until the set of terms is closed under $\partial^2$, gives:
\begin{align}
    \nonumber S^{(2)} &= \frac{\beta^3}{72000}\Big[-4467\Psi^{(1)} + 1305\Psi^{(3)} -990\Tilde\Psi^{(2,1)} -300 \Tilde\Psi^{(1,2)}_d + 200\Tilde\Psi^{(1,1,1)}_{branch} \\
    & \qquad + \nonumber 225\Tilde\Psi^{(1,1,1)}_{chain} + 246\Tilde\Psi^{(1,2f)} -210\Tilde\Psi^{(1,1f,1)} +35\Tilde\Psi^{(1,1f,1f)} \Big] 
\end{align}
where,
\begin{align}
    \nonumber &\Psi^{(1)} \equiv \sum_x \sum_{\mu} {s_{x}} \cdot {s_{x+\mu}} \quad \quad  
    \nonumber \Psi^{(3)} \equiv \sum_x \sum_{\mu,\nu,\lambda} {s_{x}}\cdot{s_{x+\mu+\nu+\lambda}} \quad \quad 
    \nonumber \Tilde\Psi^{(2,1)} \equiv \sum_x \sum_{\mu,\nu,\lambda} ({s_{x}} \cdot {s_{x+\mu+\nu}})({s_{x}} \cdot {s_{x+\lambda}}) \\
    \nonumber &\Tilde\Psi^{(1,2)}_{disc} \equiv \sum_x \sum_{\mu,\nu,\lambda} ( {s_{x}} \cdot {s_{x+\mu}})({s_{x+\nu}} \cdot {s_{x+\lambda}}) \quad \quad
    \nonumber \Tilde\Psi^{(1,1,1)}_{branch} \equiv \sum_x \sum_{\mu,\nu,\lambda} ({s_{x}} \cdot {s_{x+\mu}})({s_{x}}\cdot{s_{x+\nu}})({s_{x}}\cdot{s_{x+\lambda}})\\
    \nonumber &\Tilde\Psi^{(1,1,1)}_{chain} \equiv \sum_x \sum_{\mu,\nu,\lambda} ({s_{x}}\cdot{s_{x+\mu}})({s_{x+\mu}}\cdot{s_{x+\mu+\nu}})({s_{x+\mu+\nu}}\cdot{s_{x+\mu+\nu+\lambda}}) \quad \quad
    \nonumber \Tilde\Psi^{(1,2f)} \equiv \sum_x \sum_{\mu,\nu} ({s_{x}}\cdot{s_{x+\mu}})({s_{x}}\cdot{s_{x+\mu+\nu}})\\
    \nonumber &\Tilde\Psi^{(1,1f,1)} \equiv \sum_x \sum_{\mu,\nu} ({s_{x}}\cdot{s_{x+\mu}}^2)({s_{x}}\cdot{s_{x+\nu}})\quad \quad
    \nonumber \Tilde\Psi^{(1,1f,1f)} \equiv \sum_x \sum_{\mu} ({s_{x}}\cdot{s_{x+\mu}})^3
\end{align}

This process may be continued to arbitrary order, however, the set of terms grows rapidly. The 3rd order has 40 unique types of terms. The 4th order has 176 unique types of terms. Therefore, the procedure becomes quickly intractable as it proceeds to higher orders. A Rust and Python implementation has been made to calculate the terms to arbitrary order (and cross-check), but these terms were not implemented in our simulation~\footnote{  See {\tt https://github.com/cdchamness/O3OrderExpansion}}. 

\section{Numerical Tests}\vspace{-0.2cm}
Using the approximate solution detailed above defines a flow that approximately maps trivial variables to those following the distribution of the 2D-O(3) model. By applying this map on uniformly distributed fields of unit vectors, it is possible to generate samples that are close to the target distribution. Using Eq.~\eqref{SimpleTrivFlow} we measure the deviation from the true distribution. A perfect flow would always result in the LHS being equal to the RHS. Therefore the error on the map may be determined by the variance of the differences between the approximate distribution and the target distribution. A variance of the difference is used as any overall shift may be absorbed into the definition of the action and has no effect on the distribution.
\newcommand{\Err}{\sigma_{\Delta S}}
\begin{equation}
    \sigma_{\Delta S} = Var(\Delta S)^{1/2} \equiv Var\Big(\mathcal{S}(s) - \Big[ S_0(\mathcal{F}(s)) - \ln \mathcal{J}(s) \Big]\Big)^{1/2}
\end{equation}

\begin{figure}[t]
\centering
  \includegraphics[width=.45\linewidth]{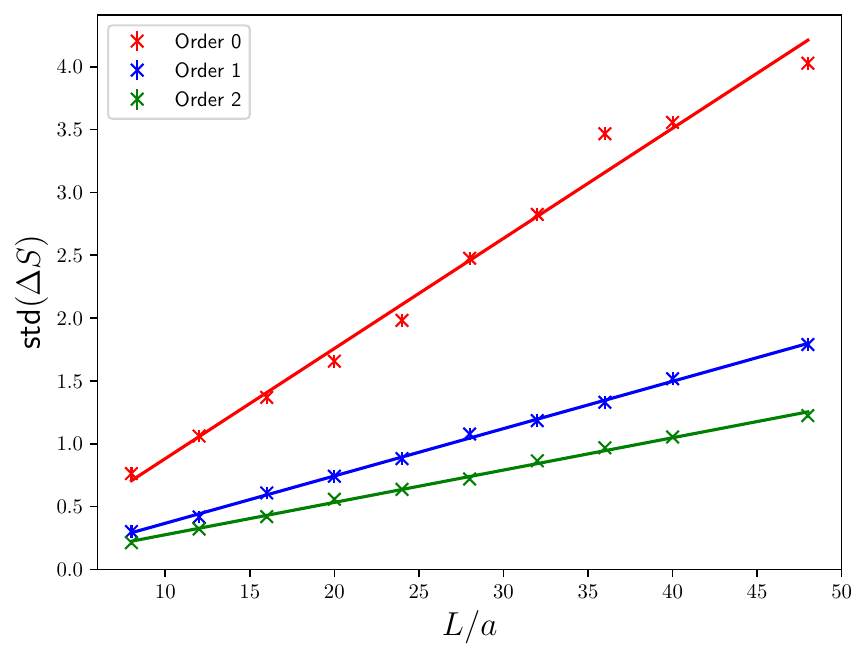}
  \includegraphics[width=.45\linewidth]{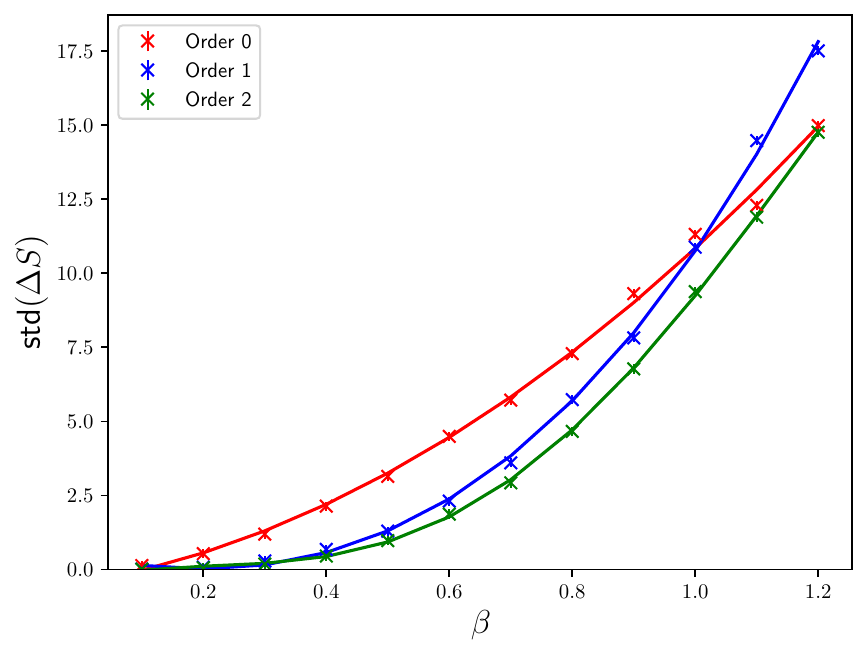}
  \captionof{figure}{The standard deviation of $\Delta S$ as a function of the Lattice Size (left) and coupling $\beta$ (right). The lines are fitted to the data as described in the text. 
  \label{fig:test1}}
\end{figure}

To investigate the accuracy of the flow, simulations were performed to apply the flow on trivial fields. For each measurement, we considered 400 configurations, where the flow was applied via a fixed step size of $\delta t=0.05$. Jackknife sampling was used to determine the errors of the variances. We computed the scaling of the $\Err$ as a function of the lattice volume $V$ and the coupling $\beta$ of the target action. 
We observe linear scaling of $\Err$ with the lattice side length $L=V^{1/2}$ as shown in Figure \ref{fig:test1}. This test was performed at fixed $\beta = 0.5$. As higher order terms are included $\Err$ is reduced  for all lattice sizes $L$. Furthermore, at fixed order in the flow time expansion, the scaling of $\Err$ is approximately linear with the lattice size $L$ in lattice units. The linear scaling of $\Err$ with lattice size suggests that the quality of the approximation of the target distribution via the approximate Trivializing map will drop exponentially with $L$.
In Figure \ref{fig:test1} we consider the scaling  of $\Err$ vs. $\beta$ while keeping the lattice size fixed, $L=36$. 
Our solution is an expansion in $0 \le t \le 1$ however, at every order, an additional factor of $\beta$ is obtained through the contraction with $\mathcal{S}$. For that reason the truncation to a fixed order results in polynomial corrections in $\beta$ at small $\beta$ ($\Err ~ \beta^2$(Ord. 0), $\Err ~ \beta^3$(Ord. 1),  $\Err ~ \beta^4$(Ord. 2)).  
From our current experiments, it is not clear what the behavior of the approximation is as one approaches criticality at large $\beta$ and volumes.

We may improve the approximation by adding a subset of known terms to approximate higher-order solutions with little increase in computational cost. By noticing that $\Psi^{(1)}$ appears in both the 0th order and 2nd order solutions it is then possible to know that all terms that appeared in the 1st order solution will appear in the 3rd order solution. This is because the process for determining the initial set of each order is the same, specifically $\sum_x \partial_x \mathcal{S} \; \partial_x S^{(n-1)}$. This also works with respect to the 2nd-order and 4th-order solutions. Therefore it is possible to determine which terms appear in "even" order solutions and terms that appear in "odd" order solutions.
\begin{align}
    S^{(2n)} = \beta^{2n+1}\sum_a \gamma^{(2n)}_a \Psi_{even}^a \quad \quad 
    S^{(2n+1)} = \beta^{2n+2}\sum_a \gamma^{(2n+1)}_a \Psi_{odd}^a
\end{align}
By non-perturbative tuning of the $\{\gamma_\alpha^{(n)}\}$, it might be possible to increase performance further with little cost to performing the flow, as these terms would already be computed for previous orders. 

\section{Conclusions}\vspace{-0.2cm}
In this contribution, we present an analytic construction of a trivializing flow for the 2D-O(3) sigma model. The construction is based on a gradient flow on the $\mathcal S^2$ manifold, whose generating functional is determined order-by-order in the small flow-time expansion. This perturbative expansion is algorithmic and has been programmed in principle to all orders. However, it is impractical beyond the first few orders. We numerically tested the efficiency of the approximation and found evidence that the computational cost of the approximation scales exponentially with the lattice size. We intend to investigate this issue further both numerically and analytically. If such behavior persists it is doubtful that trivializing maps constructed perturbatively or non-perturbatively can eliminate critical showing down which occurs close to criticality where large correlation lengths mandate large lattices in lattice units.    

{\bf Acknowledgements:}
This work was supported by the U.S. DOE Grant DE-FG02-04ER41302 and by the US DOE Contract No. DE-AC05-06OR23177. We thank Balin Armstrong for discussions related to his work on trivializing maps of the O(2) sigma model. 

\bibliography{biblio}

\end{document}